\documentclass[superscriptaddress,showpacs,reprint]{revtex4-1}
\usepackage{graphicx}
\usepackage{dcolumn}
\usepackage{bm} 
\usepackage{amsmath}
\usepackage{color}
\definecolor{dark-red}{rgb}{0.9,0.15,0.15}
\definecolor{dark-blue}{rgb}{0.15,0.15,0.4}
\definecolor{medium-blue}{rgb}{0,0,0.5}
\usepackage{hyperref}
\hypersetup{
	colorlinks=true,
	citecolor=blue,
	linkcolor=blue,
	filecolor=blue,      
	urlcolor=blue,
}
\usepackage[most]{tcolorbox}

\begin{document}

\title{Spin-flop quasi metamagnetic, anisotropic magnetic, and electrical transport behavior of Ho substituted kagome magnet ErMn$_6$Sn$_6$}

\author{Jacob Casey}
\affiliation{Department of Physics, SUNY Buffalo State University, Buffalo, NY 14222, United States}

\author{S. Shanmukharao Samatham}
\email{shanmukharao\_physics@cbit.ac.in}

\affiliation{Department of Physics, Chaitanya Bharathi Institute of Technology, Gandipet, Hyderabad 500 075, India}

\author{Christopher Burgio}
\affiliation{Department of Physics, SUNY Buffalo State University, Buffalo, NY 14222, United States}

\author{Noah Kramer}
\affiliation{Department of Physics, SUNY Buffalo State University, Buffalo, NY 14222, United States}
\affiliation{Engineering Technology, SUNY Buffalo State University, University, NY 14222, United States}

\author{Asraf Sawon}
\affiliation{Department of Physics, SUNY Buffalo State University, Buffalo, NY 14222, United States}
\affiliation{Engineering Technology, SUNY Buffalo State University, University, NY 14222, United States}

\author{Jamaal Huff}
\affiliation{Department of Physics, SUNY Buffalo State University, Buffalo, NY 14222, United States}
\affiliation{Engineering Technology, SUNY Buffalo State University, University, NY 14222, United States}

\author{Arjun K. Pathak}
\email{pathakak@buffalostate.edu}
\thanks{Current address: GE Research, Niskayuna, NY 12309, USA}
\affiliation{Department of Physics, SUNY Buffalo State University, Buffalo, NY 14222, United States}

\begin{abstract}
We report on the magnetic and electrical properties of a (Mn$_3$Sn)$_2$ triangular network kagome structured high quality Ho substituted ErMn$_6$Sn$_6$ single-crystal sample by magneto-transport measurements. Er$_{0.5}$Ho$_{0.5}$Mn$_6$Sn$_6$ orders antiferromagnetically at N\'{e}el temperature $T_\mathrm{N} \sim$ 350 K followed by a ferrimagnetic (FiM) transition at $T_\mathrm{C} \sim$ 114 K and spin-orientation transition at $T_\mathrm{t} \sim$ 20 K. The field-manifestations of these magnetic phases in the \textit{ab}-basal plane and along the \textit{c}-axis are illustrated through temperature-field \textit{T-H} phase diagrams. In \textit{H}$\parallel$\textit{c}, narrow hysteresis between spin reorientation and field-induced FiM phases below $T_\mathrm{t}$, enhanced/strengthened FiM phase below $T_\mathrm{C}$ and stemming of FiM phase out of strongly coexisting AFM and FiM phases below $T_\mathrm{N}$ through a non-meta-magnetic transition are confirmed to arise from strong R-Mn sublattices interaction. In contrast, \textit{H}$\parallel$\textit{ab}-plane, between $T_\mathrm{N}$ and $T_\mathrm{C}$, individually contributing R-Mn sublattices with weak antiferromagnetic interactions undergo a field-induced spin-flop quasi-metamagnetic transition to FiM state. The temperature dependent electrical resistivity suggests metallic nature with Fermi liquid behavior at low temperatures. Essentially, the current study stimulates interest to investigate the magnetic and electrical properties of mixed rare-earth layered kagome magnetic metals for possible novel and exotic behavior.
\end{abstract}

\date{\today}

\maketitle

\section{Introduction}

In recent times, condensed matter research has been evidencing an intensified search and investigation of materials with quantum phenomena driven exotic properties pertaining to their potential role in the fundamental exploration of physical phenomenon and technological applications. Such an exotic behavior is thought to be feasible when topology is effectively correlated and coupled with the magnetic field/magnetism. Topological insulators with unidirectional edge-conducting paths are expected to pave a new path for low-loss electrical conduction devices technologically. Contemporarily, transition metal based monosilcides, EuPtSi, Cu$_2$OSeO$_3$, $\beta$-Mn-type Co-Zn-Mn alloys, and some of the Heusler alloys are being studied actively for low-energy level skyrmion phase in view of fast data transfer of spintronics and large storage capacity. While the anomalous Hall effect is commonly observed in ferromagnetically ordered materials due to symmetry (time-reversal) breaking, the topological Hall effect finds its routes in strong spin-orbit coupling leading to large Berry curvature. In general, the geometry of (chemical/magnetic) lattice and the crystal symmetry mainly govern the electronic properties such as density of states, Fermi level etc of a crystal. Recently, kagome lattices, with corner sharing triangular networks of transition metals within the plane (two-dimensional) have been reported to emerge as potential facilitators for correlated and topological phenomena \cite{Kagome_topological_General,EMS_PhysRevLett_2021}. Such a peculiar geometry of transition metals in kagome lattice leads to a linear (conduction-valence) band-crossing point resulting in Dirac Fermions, geometrical-frustrated driven spin liquid phenomena and noncollinear anti-parallel spin configuration. To exemplify, Dirac Fermions in Fe$_3$Sn$_2$, a ferromagnet \cite{Fe3Sn2_kagome_dirac_2019} and FeSn, an antiferromagnet \cite{FeSn_1_PhysRevMaterials_2019,FeSn_2_Sci_Rep_2022,FeSn_3_Commu_Phys_2021}, Co$_3$Sn$_2$S$_2$ with Weyl Fermions \cite{Co3Sn2S2_topological_2021} and Mn$_3$Sn with noncollinear antiferromagnetism \cite{Mn3Sn_1_AFM_2017,Mn3Sn_2_AFM_2019}. Chern-gapped Dirac fermions in ferromagnetic TbMn$_6$Sn$_6$ \cite{TMS_topological_2022}. The kagome compounds are predicted to host Chern gapped Dirac Fermions with strong spin-orbit coupling and parallel spin configuration in the out-of-plane \cite{Quantum_Hall_PhysRevLett_2011,Quantum_Hall_PhysRevLett_2015}.

Belonging to such kagome family, the rare-earth-transition metal based intermetallic alloys with chemical formula RMn$_6$Sn$_6$ (R is rare-earth element Gd, Er, Ho, Tb, Tm, Lu) with hexagonal structure exhibit (Mn$_3$Sn)$_2$ kagome network. In addition, a long-range magnetic order with strong spin-orbit coupling in these kagome-magnets make them interesting to investigate the quantum-phenomena led properties. Neutron diffraction study of RMn$_6$Sn$_6$ by Malaman \textit{et al.}, \cite{RMS_Malaman_JMMM_1999} reveals that i) high ordering temperature is because of strong Mn-Mn intra-sublattice ferromagnetic exchange interactions and ii) collinear ferrimagnetic ordering below 75 K is because of the simultaneous ordering of R-Mn sublattices with antiferromagnetic coupling in ErMn$_6$Sn$_6$. Magnetic study of flux-grown single crystals of RMn$_6$Sn$_6$ by Clatterbuck \textit{et al.}, \cite{Structure_Clatterbuck_JMMM_1999} report ferrimagnetic ordering for Tb-and Ho compounds respectively at 450 and 410 K; while R = Er, Lu, Tm are antiferromagnetic in the absence of magnetic field. A complex magnetic structure was projected for ErMn$_6$Sn$_6$ with an antiferromagnetic Er and Mn-sublattices at 65 K and 355 K respectively in zero-field. Eventually, Er and Mn sublattices separately turn in to parallel-spin configuration in high fields, however, with an anti-parallel correlation between the sublattices. It is emphasized that the interaction between Mn-Mn atoms is sensitive to the magnitude of lattice parameters, resulting in the complex magnetic behavior of ErMn$_6$Sn$_6$.

Given a complex magnetic behavior resulting from antiferromagnetically coupled usual competing ferromagnetic orderings of R and Mn sublattices, nuclear magnetic resonance (NMR) study by Shimizu and Hori \cite{EMS_NMR_Shimizu} suggested magnetic moments of 8.9 and 2.2 $\mu_\mathrm{B}$ for Er and Mn respectively in the ferrimagnetic state of ErMn$_6$Sn$_6$ at 4.2 K. In addition, high-field (up to 500 kOe) magnetization isotherms show multiple field-induced transitions in the basal plane \cite{EMS_High_Field_Suga_JALCOM}. Almost matching critical fields of free-powder sample with that of single crystal has led to the conclusion that field-induced metamagnetic transitions in ErMn$_6$Sn$_6$ are exchange interaction dominated instead of resulting from the anisotropy \cite{EMS_Meta_Hu_1999}, however with weak Mn-Mn coupling \cite{EMS_Magnetoelastic_Yazdi_2012}. Nevertheless, an experimental and theoretical investigation of single crystal ErMn$_6$Sn$_6$ (with antiferromagnetic $T_\mathrm{N}$ = 345 K and ferrimagnetic $T_\mathrm{C}$ = 68 K) by Dhakal \textit{et al.}, \cite{EMS_Pathak_Dhakal_PhysRevB}, our recent paper, report direct observation of large anisotropic anomalous and topological Hall signatures arising from multi-orbital band structure at the Fermi level. Another compound with kagome network magnet HoMn$_6$Sn$_6$ exhibits ferrimagnetism below $T_\mathrm{C}$ = 376 K followed by a spin-reorientation at $T_\mathrm{t}$ = 200 K \cite{HMS_Kabir_PhysRevMater_2022} with topological and anomalous Hall effect properties \cite{HMS_Kabir_PhysRevMater_2022,HMS_JALCOM_2022}. Contrasting magnetic behaviors above and below room temperature, large variation of $\Delta T_\mathrm{order} \sim$ 100 K (the difference between two magnetic ordering temperatures) and exhibition of topological transport properties linked with magnetic manifestations in HoMn$_6$Sn$_6$ and ErMn$_6$Sn$_6$ have stimulated interest to investigate the magnetic behavior of single crystal Ho-half substituted ErMn$_6$Sn$_6$. Further, it would be interesting to know the manifestation of magnetic and physical properties of Er$_{0.5}$Ho$_{0.5}$Mn$_6$Sn$_6$ given the dissimilar magnetic characters of free ion Er (Kramer-doubly degenerate states with odd number of 4\textit{f} electrons) and free ion Ho (non-Kramer singlets with even number of 4\textit{f} electrons) because of zero-field splitting. In addition, in a kagome layered structured materials, as the magnetic interactions between R (rare-earth) and Mn atoms play a crucial role, the effect of magnetic field on these crystals is expected to modify the magnetic ground states by removing the degeneracy.

Our work presents the investigation of Er$_{0.5}$Ho$_{0.5}$Mn$_6$Sn$_6$ single crystal making use of structural and magnetization characterizations. It exhibits dual magnetic transitions ($T_\mathrm{N} \sim$ 350 K and $T_\mathrm{C} \sim$ 114 K). The temperature-magnetic field \textit{T-H} phase diagrams demonstrate the manifestation of anisotropic magnetic properties in the \textit{ab}-plane and along \textit{c}-axis. In the \textit{ab}-plane, a field-induced ferrimagnetic phase, without first-order phase transition, out of zero-field coexisting ferri and antiferromagnetic phases is realized. Such coexistence caused a field-induced narrow hysteresis at low temperatures. Along the \textit{c}-axis, a field-induced quasi-metamagnetic non-first-order antiferromagnetic to ferrimagnetic transition is evidenced. The electrical resistivity suggests metallic nature with Fermi liquid behavior while negative but small magnetoresistance arises from the quenching of spin-disorder scattering in external fields.

\begin{figure}
	\centering
	\includegraphics[width=\linewidth,height=\linewidth]{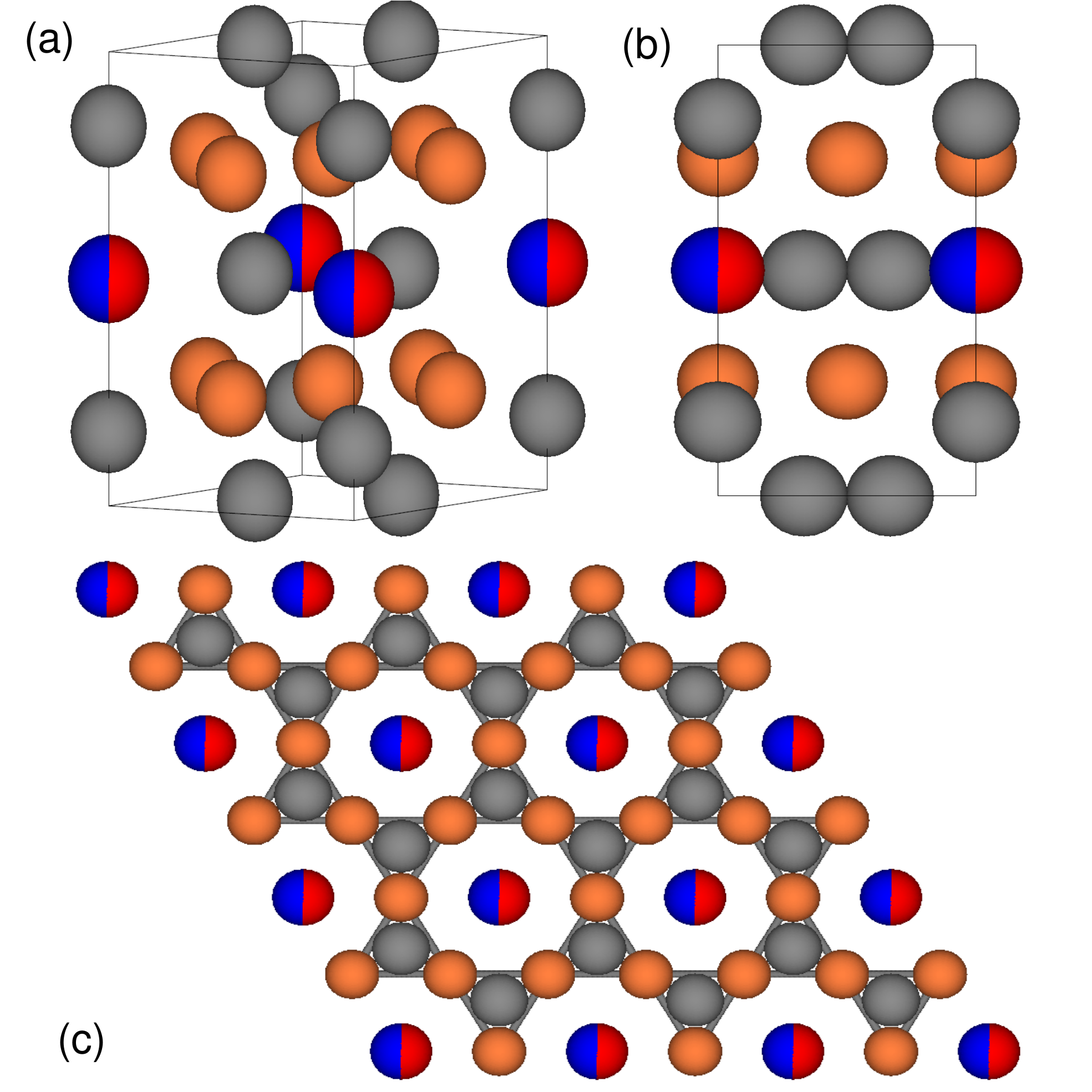}
	\caption{Crystallographic information of Er$_{0.5}$Ho$_{0.5}$Sn$_6$Mn$_6$. Color codes for atoms: Red-Er, Blue-Ho, Orange-Mn and Grey-Sn. (a) a three-dimensional view of the crystal structure in a unit cell, showing the Mn atoms symmetrically equidistant from R-atoms. (b) A layered structure with (R-Sn\_1)-Mn-Sn\_3-Sn\_2 stacks along the \textit{c}-axis. (c) corner sharing (Mn$_3$Sn)$_2$ triangular network.}
	\label{fig:xrd_structure}
\end{figure}

\section{Experimental Methods}

Single crystals of Ho substituted ErMn$_{6}$Sn$_{6}$ were grown by flux method using the protocol described in Refs. \cite{EMS_Pathak_Dhakal_PhysRevB,HMS_Kabir_PhysRevMater_2022}. As grown crystals were properly polished and cleaned by chemical etching before any physical property measurements. The composition and crystal structure of Er$_{0.5}$Ho$_{0.5}$Mn$_6$Sn$_6$ (MgFe$_6$Ge$_6$-type hexagonal) were verified by Energy Dispersive X-ray Analysis (EDAX) and Laue x-ray diffraction as given in the Supplementary Material \cite{Supplement}. The magnetization measurements were performed using a Quantum Design Physical Property Measurement System (PPMS) equipped with a vibrating sample magnetometer (VSM). The measurements were performed both along \textit{c}-axis and in the \textit{ab}-plane i.e. perpendicular to \textit{c}-axis. The temperature dependent magnetization $M(T)$ was measured in zero-field cooling (ZFC) and field-cooled warming (FCW) processes. In the ZFC process, a required magnetic field of measurement was applied at 2 K after cooling the crystal from 395 K in the absence of externally applied magnetic fields and consequently the magnetization was recorded (from 2 K to 395 K) during warming. In the FCW process, the magnetization isotherms as a function of magnetic field $M(H)$ was collected during warming from 2 K to 395 K, after initially cooling the crystal down to 2 K in the presence of magnetic fields. An isothermal magnetization as a function of magnetic field \textit{H} was recorded in ZFC process. A maximum of \textit{H} = 90 kOe was applied to record $M(H)$. The electrical resistivity of Er$_{0.5}$Ho$_{0.5}$Mn$_6$Sn$_6$ was measured using a standard dc-four probe method, as a function of temperature and magnetic field, using PPMS at temperature between 2 to 390 K and magnetic field up to 90 kOe.

\section{Results}

Figure \ref{fig:xrd_structure}(a) is a three dimensional (3D) view of the unit cell drawn using visualization for electronic and structural analysis (VESTA) software \cite{VESTA}. The atomic/Wyckoff positions of the elements are as follows: Er and Ho share 1(a) (0, 0, 0) position and Mn occupies 6(i) (1/2, 0, \textit{z}) with \textit{z} = 0.74712. The element Sn occupies three different positions namely Sn\_1 2(c) (1/3, 2/3, 0), Sn\_2 2(d) (1/3, 2/3, 1/2) and Sn\_3 2(e) (0, 0, $z \sim$ 0.83641) \cite{Structure_Clatterbuck_JMMM_1999}. It shows that Er/Ho is located at four edges (4$\times$1/4 = 1), Mn: two at the center and 6 on the faces (2$\times$1+8$\times$1/2 = 6) and Sn: 8 on the edges, two each on the opposite faces and two at the center (8$\times$1/4+4$\times$1/2+2$\times$1 = 6). Figure \ref{fig:xrd_structure}(b) shows the consecutive layers Sn-Sn-Mn-(R-Mn)-Mn-Sn-Sn with a view along \textit{a}-axis. Figure \ref{fig:xrd_structure}(c) presents a view along \textit{c}-axis, showing corner sharing (Mn$_3$Sn)$_2$ triangular network forming a kagome lattice.

\begin{figure}
	\centering
	\includegraphics[width=\linewidth,height=1.5\linewidth]{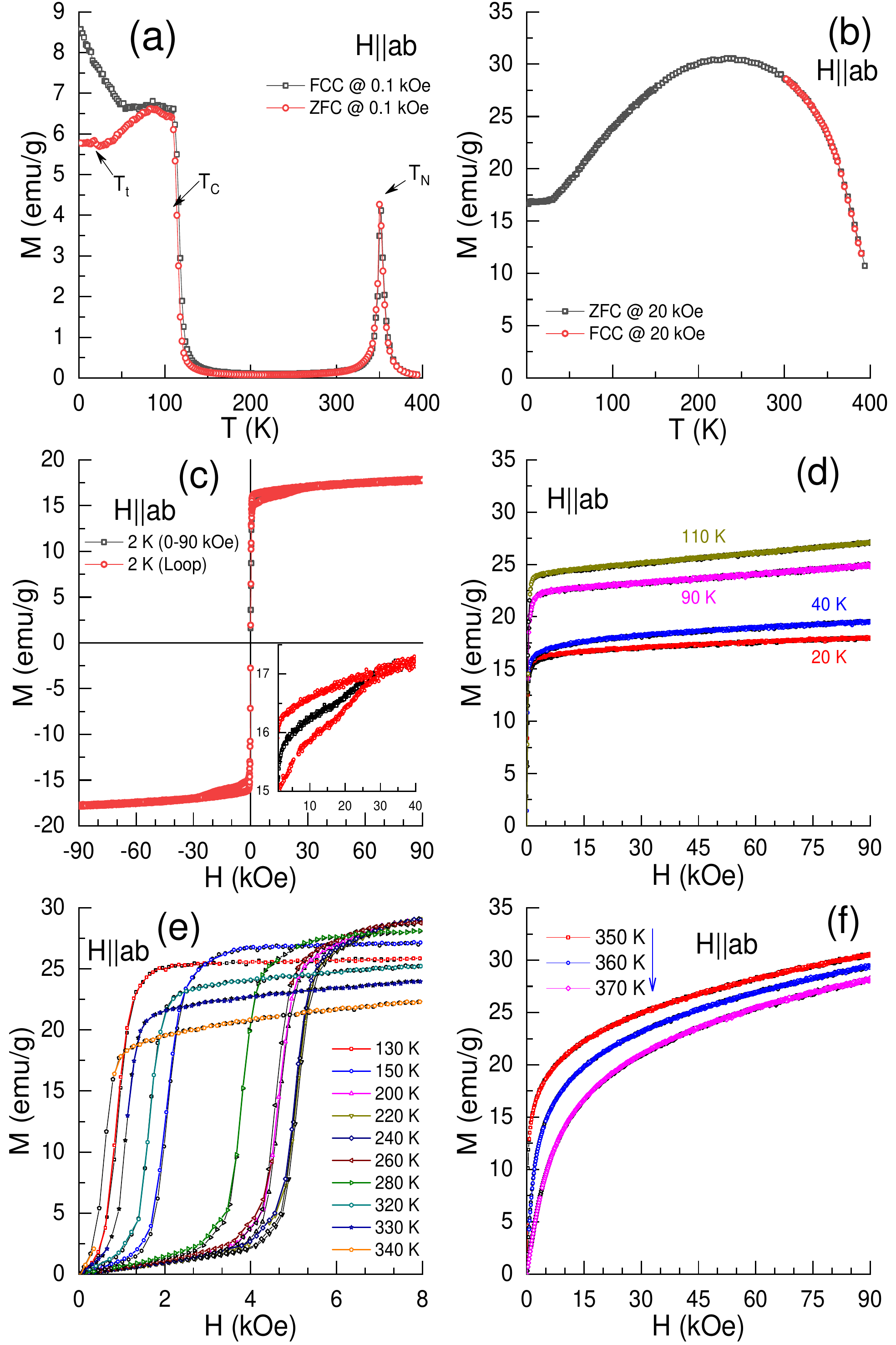}
	\caption{Magnetization data of Er$_{0.5}$Ho$_{0.5}$Mn$_6$Sn$_6$ single crystal measured \textit{H}$\parallel$\textit{ab}: ZFC, FCC \textit{M-T} curves (a) in 0.1 kOe, shows antiferromagnetic transition at 350 K followed by a ferrimagnetic transition around 110 K and a spin-reorientation-type transition below 20 K and a bifurcation between ZFC and FCC curves and (b) in 20 kOe, where antiferromagnetic transition is smeared out in fields. An isothermal magnetization vs. magnetic field \textit{M-H} curves up to \textit{H} = 90 kOe (c) at 2 K (five quadrants, showing reversible nature with no-hysteresis). Inset: The virgin curve (0$\rightarrow$90 kOe) shows a tiny field-induced transition around 20 kOe, (d) in the ferrimagnetic region i.e., $T_\mathrm{t} \le T \le T_\mathrm{C}$, (e) in the region bounded by $T_\mathrm{C}$ and $T_\mathrm{N}$, showing the manifestation of a field-induced AFM to FiM crossover and (f) above $T_\mathrm{N}$ i.e. in the paramagnetic state with no saturation-like and reversible nature.}
	\label{fig:mag_along_c}
\end{figure}

Figure \ref{fig:mag_along_c}(a) shows the temperature and magnetic field dependent magnetization of Er$_{0.5}$Ho$_{0.5}$Mn$_6$Sn$_6$ when field is applied in the \textit{ab}-plane i.e. \textit{H}$\parallel$\textit{ab}, from 2-390 K. The ZFC and FCC magnetization were measured in presence of 0.1 kOe. A sharp transition around $T_\mathrm{N}$ = 350 K is designated as the paramagnetic to antiferromagnetic transition. By lowering the temperature, the crystal undergoes another magnetic transitions near 110 K, designated as $T_\mathrm{C}$, until which the crystal shows an antiferromagnetic behavior. Further reduction of temperature has resulted in another magnetic transition named spin reorientation transition $T_\mathrm{t} \sim$ 20 K. Overall, these three consecutive magnetic transitions, in principle, enable four different temperature regimes of study for Er$_{xx}$Ho$_{xx}$Mn$_6$Sn$_6$ namely I) $T < T_\mathrm{t}$, II) $T_\mathrm{t} \le T \le T_\mathrm{C}$, III) $T_\mathrm{C} \le T \le T_\mathrm{N}$ and IV) $T > T_\mathrm{N}$. Figure \ref{fig:mag_along_c}(b) shows \textit{M-T} data measured in \textit{H} = 20 kOe from 2-390 K. The data shows an indifference between ZFC and FCC with a single peak around 230 K, suggesting field-conversion of antiferromagnetic to ferrimagnetic state. It can be noted that the magnetic field suppresses $T_\mathrm{N}$ and enhances $T_\mathrm{C}$ and these two transitions merge in relatively high magnetic fields (of the order of 20 kOe). Figure \ref{fig:mag_along_c}(c) shows an isothermal magnetization \textit{M-H} curve recorded at 2 K (i.e., in the temperature region-I) and in five quadrants: 0 $\rightarrow$ 90 kOe (virgin curve), 90 $\rightarrow$ 0 kOe, 0 $\rightarrow$ -90 kOe, -90 $\rightarrow$ 0 kOe and 0 $\rightarrow$ 90 kOe. It is evident that the magnetization of Er$_{0.5}$Ho$_{0.5}$Mn$_6$Sn$_6$ is reversible under field cycles with a sharp non-linear approach as \textit{H}$\rightarrow$ 0 kOe. In addition, the hysteresis loops open and close at the origin (\textit{H} = 0, \textit{M} = 0) with zero coercive magnetic field. The symmetric hysteresis loop indicates with no coercive magnetic field rules out the possibility of spontaneous exchange bias in the crystal along the \textit{c}-axis. Figure \ref{fig:mag_along_c}(d) depicts the magnetization and demagnetization \textit{M-H} isotherms, similar though, in the temperature region-II. Commonly in all the isotherms, the magnetization sharply rises until technical saturation magnetic field $H_\mathrm{TS}$ and thereafter it increases linearly with the applied field. This observation points out the fact that the magnetic field converts the zero-field ferrimagnetic to a high-field ferrimagnetic-saturation like state above $H_\mathrm{TS}$ in the \textit{T}-region-II. In particular, it is noteworthy to observe that the saturation magnetic moment $M_\mathrm{s}$, the magnetization intercept of the linear extrapolation of high-field magnetization down to zero magnetic field, increases with temperature, in this \textit{T}-region.

Figure \ref{fig:mag_along_c}(e) shows the magnetization isotherms in the temperature region-III from 130 to 340 K bounded by $T_\mathrm{C}$ and $T_\mathrm{N}$. This is the region where low-field magnetic ground state is antiferromagnetic in the \textit{ab}-plane. \textit{M-H} at 130 K exhibits a critical magnetic field $H_\mathrm{cr}$ (later named as $H_\mathrm{meta}$) followed by a technical saturation field $H_\mathrm{TS}$. $H_\mathrm{meta}$ is the critical field at which the antiferromagnetic spin alignment starts to transform into ferrimagnetic spin configuration while $H_\mathrm{TS}$ is the critical field above which Er$_{0.5}$Ho$_{0.5}$Mn$_6$Sn$_6$ exhibits ferrimagnetic linearly-saturating behavior. The data reveal that $H_\mathrm{meta}$, $H_\mathrm{TS}$ and the spontaneous magnetic moment increase with temperature by showing a maximum value at 220 K. Further increasing the temperature above 260 K i.e., while approaching $T_\mathrm{N}$, these values decrease. Figure \ref{fig:mag_along_c}(f) shows the magnetization isotherms in the \textit{T}-region-IV i.e. the paramagnetic state. \textit{M}-\textit{H} curves above $T_\mathrm{N}$, in both \textit{H}$\parallel$\textit{c} and \textit{H}$\parallel$\textit{ab} configurations, are not found to be linear throughout the applied field range. However, a smooth increase of \textit{M}-\textit{H} with a strong curvature, without field-induced transition and an almost linear behavior at larger fields indicate polarization of magnetic moments by magnetic field just above $T_\mathrm{N}$. This infers to the correlated paramagnetic state with dominant magnetic correlations over the randomizing thermal fluctuations as is also evident from the large positive $\theta_\mathrm{CW}$. However, the compound is expected to show paramagnetic behavior with linear \textit{M}-\textit{H} at elevated temperatures ($T>>T_\mathrm{N}$).

\begin{figure}
	\centering
	\includegraphics[width=\linewidth,height=1.5\linewidth]{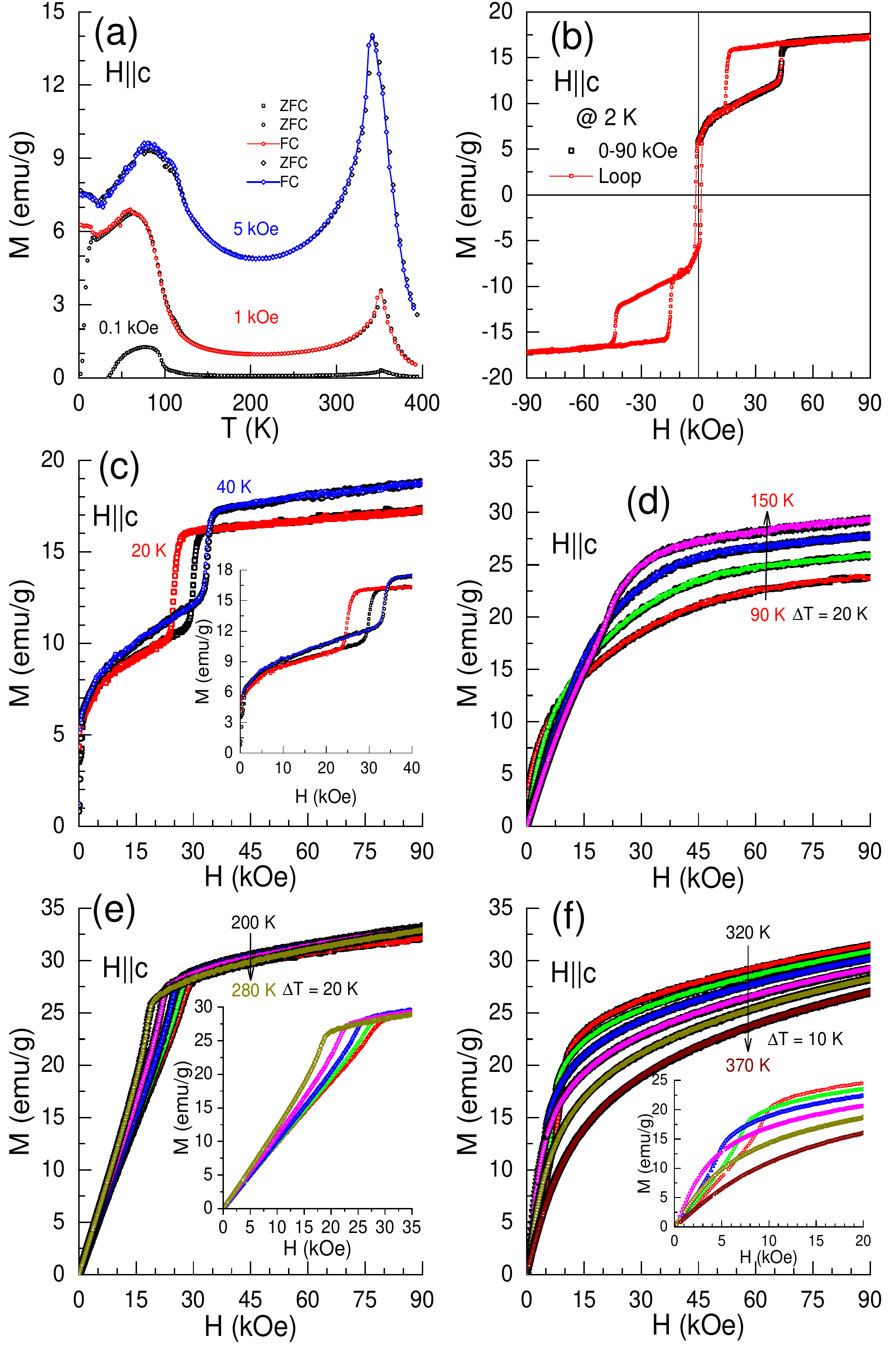}
	\caption{Magnetization data of Er$_{0.5}$Ho$_{0.5}$Mn$_6$Sn$_6$ single crystal when \textit{H}$\parallel$\textit{c}: \textit{M-T} curves (a) in 0.1, 1 and 5 kOe, showing transitions at $T_\mathrm{N}$ = 350 K, $T_\mathrm{C}$ = 110 K and $T_\mathrm{t} \sim$ 23 K. An isothermal \textit{M-H} curves up to \textit{H} = 90 kOe (b) at 2 K in five quadrants, showing a small hysteresis between a field-induced FiM state, (c) at 20 and 40 K. Inset: vanishing hysteresis above $T_\mathrm{t}$, (d) collected at temperature range of 90 K $\le T \le$ 150 K, showing a non-linear growth of \textit{M} with \textit{H} with an improved linear-saturation as \textit{T} crosses $T_\mathrm{C}$, (e) in 200 K $\le$ T $\le$ 280 K where a field-sustaining linear \textit{M-H} up to a range of 30 kOe before reaching FiM-like saturation in high fields, indicates the coexistence of AFM and FiM states. Inset: An enlarged view (up to 35 kOe) showing decreasing critical field above 200 K.and (f) near and above $T_\mathrm{N}$. The critical field decreases and eventually becomes zero as \textit{T} approaches $T_\mathrm{N}$.}
	\label{fig:mag_along_ab}
\end{figure}

Figure \ref{fig:mag_along_ab} shows the temperature and field dependent magnetization data measured when \textit{H} is applied in the \textit{c}-axis i.e., \textit{H}$\parallel$\textit{c}. Figure \ref{fig:mag_along_ab}(a) shows \textit{M-T} curves in \textit{H} = 0.1, 1 and 5 kOe. \textit{M-T} in 0.1 kOe exhibits three consecutive transitions at $T_\mathrm{N} \sim$ 350 K, $T_\mathrm{C}\sim$ 110 K and $T_\mathrm{t}\sim$ 23 K, akin to that of noticed in \textit{ab}-plane. However, the magnitude of magnetic moment is lower in the \textit{c}-axis. In addition, the transitions are sharp in the \textit{ab}-plane while broadened along \textit{c}-axis. Such a broadened transition indicates that the magnetic entropy (exchange interactions) is distributed across the transition. It is evident from the figure that the observed bifurcation between ZFC and FCC magnetization curves in \textit{H} = 1 kOe disappears in \textit{H} = 5 kOe, indicating a delicate nature of the magnetic state below $T_\mathrm{t} \sim$ 23 K. Figure \ref{fig:mag_along_ab}(b) shows the magnetization isotherm at 2 K in five quadrants. The isotherms are reversible with almost zero coercive magnetic field. Initially, in the presence of low fields, the system converts a ferrimagnetic state and exhibits a field-induced hysteresis between magnetizing and demagnetizing cycles, indicating the ferrimagnetic behavior below $T_\mathrm{C}$ in the \textit{c}-axis. The diminishing width of the field-induced hysteresis and an increasing $M_\mathrm{s}$ are evident from the $M(H)$ data at 20 and 40 K in the magnetizing and demagnetizing cycles, as shown in figure \ref{fig:mag_along_ab}(c), indicating that the field-induced hysteretic nature ceases out at $T_\mathrm{C}$  and it confines to \textit{T} region-II. Unlike in the \textit{ab}-plane, in \textit{T} region-I, the crystal shows multiple field-induced transitions along with hysteresis in \textit{c}-axis, indicating a first-order-like phase transition between field-polarized spin-reorientation state and high field FiM state. Figures \ref{fig:mag_along_ab}(d) and \ref{fig:mag_along_ab}(e) depict \textit{M-H} data in the \textit{T} region-III. In this region, the metamagnetic behavior is absent. As evident from the figure \ref{fig:mag_along_ab}(d), \textit{M-H} isothermal curves of 90 and 110 K show a quasi-linear growth of magnetization in low-field and high-field regimes as well, in contrast to what is observed in the \textit{ab}-plane (refer to figure \ref{fig:mag_along_c}(d)). However, \textit{M-H} data of 130 and 150 K shows linear-growth in fields above 30 kOe with decreasing $H_\mathrm{TS}$. Deep in the \textit{T} region-III, the magnetization sees a linear increase initially and changes to an almost ferrimagnetic saturating magnetic state, as shown in the figure \ref{fig:mag_along_ab}(e) and 320, 330 and 340 K curves from figure \ref{fig:mag_along_ab}(f) (see inset for clarity). Nevertheless, the saturating field decreases with increasing temperature. Eventually, in the \textit{T}-region IV i.e., paramagnetic state, the magnetization increases smoothly, yet slightly non-linearly, without any field-induced transition. Eventually, the difference in the magnitude of magnetic moments, the nature of field-induced transitions, and varied zero-field magnetic states in the basal plane and along \textit{c}-axis confirm anisotropic magnetic properties of Er$_{0.5}$Ho$_{0.5}$Mn$_6$Sn$_6$.

\begin{figure}
	\centering
	\includegraphics[width=\linewidth,height=1.5\linewidth]{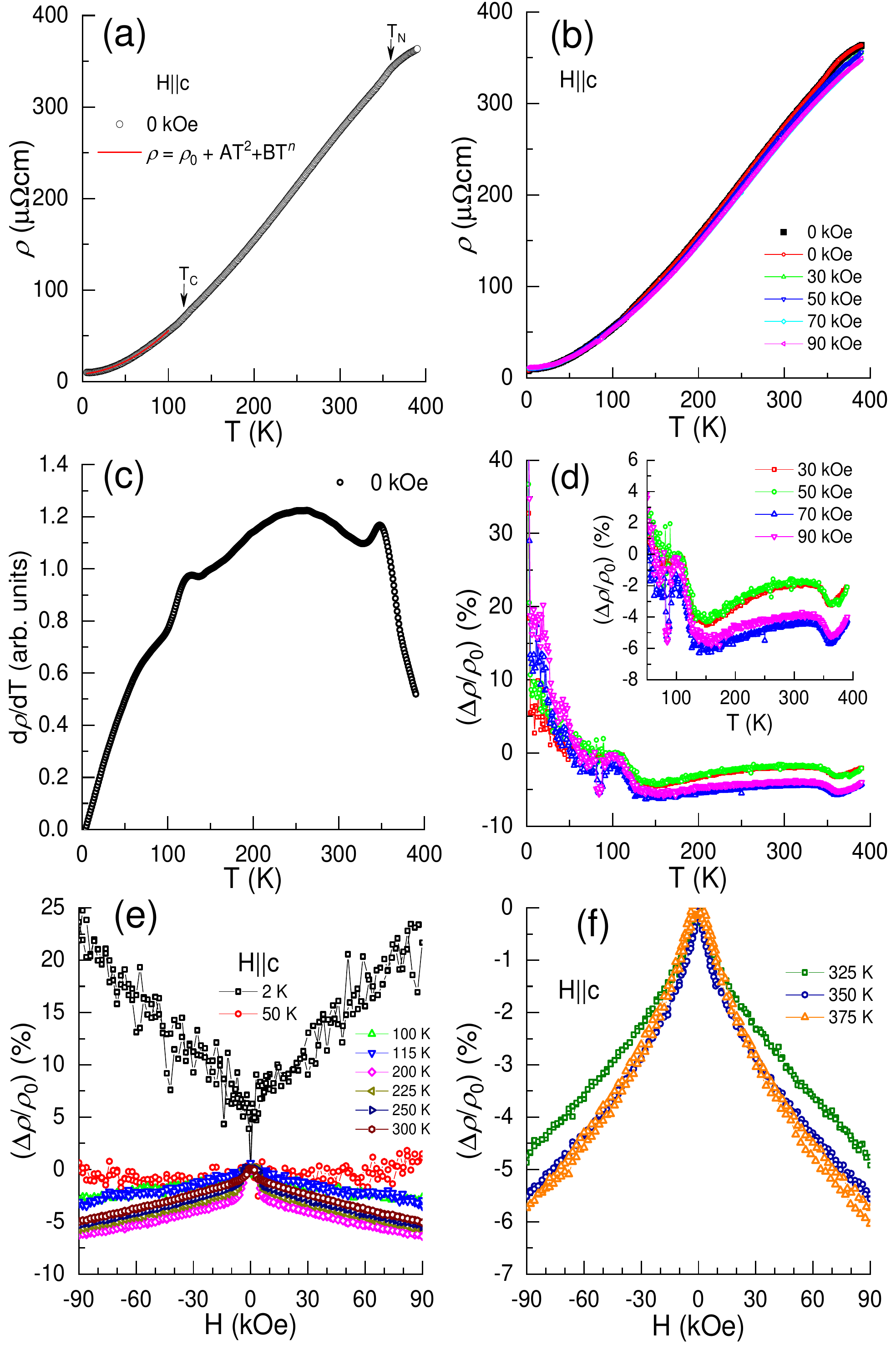}
	\caption{The temperature variation of electrical resistivity $\rho(T)$, measured \textit{H}$\parallel$\textit{c}, from 2-390 K (a) in \textit{H} = 0 kOe and (b) \textit{H} = 0, 30, 70 and 90 kOe. $\rho$ = $\rho_0+AT^2+BT^n$ fit of the data below 150 K resulted in the residual resistivity of about 8.15 $\mu\Omega$cm and \textit{n} = 3/2 indicating dominant ferrimagnetic correlations. (c) $d\rho/dT$ in $H$ = 0 kOe showing the magnetic transitions at $T_\mathrm{N}$ and $T_\mathrm{c}$, indicated by arrows in (a). (d). The temperature dependent MR = $\Delta\rho/\rho_0$, which is negative at $T_\mathrm{N}$ and $T_\mathrm{C}$ and positive below $T_\mathrm{t}$. (e-f) Magnetic field dependence of MR. It is positive below $T_\mathrm{t}$, small and negative in the other magnetic regions and paramagnetic state.}
	\label{fig:res}
\end{figure}

Figures \ref{fig:res}(a) and \ref{fig:res}(b) show the temperature dependence of electrical resistivity $\rho(T)$ from 2-390 K in \textit{H} = 0, 30, 50, 70 and 90 kOe, when \textit{H}$\parallel$\textit{c}. The resistivity data measured in heating and cooling cycles in \textit{H} = 0 kOe does not exhibit hysteresis behavior and thereby the possibility of temperature-induced first-order phase transition is discorded. The slope changes in $\rho(T)$ at the magnetic transitions $T_\mathrm{N}$ and $T_\mathrm{C}$ (indicated by arrows), identified from the temperature derivative of resistivity $d\rho/dT$ (as shown in figure \ref{fig:res}(c)) match with those observed in \textit{M-T} data. However, a slightly higher values are because of true zero-field measurements. The data below 100 K is fit using the equation $\rho-\rho_0 = AT^2+BT^n$ where $\rho_0$ is the residual resistivity, where \textit{A} is the coefficient of quadratic term, \textit{B} is the coefficient of power law dependent term and an exponent \textit{n} is found to be 3/2. The positive temperature coefficient ($d\rho/dT > 0$) confirms the metallic nature. The residual resistivity resulting from the fit is $\rho_0$ = (8.15 $\pm$ 0.1) $\mu \Omega$cm. The estimated resistivity ratio (RRR = $\rho_\mathrm{390 K}/\rho_0$) of Er$_{0.5}$Ho$_{0.5}$Mn$_6$Sn$_6$ is about 45 which is large and demonstrates the high quality of the single crystal. Furthermore, the response of electrical resistivity to the applied magnetic field is quantified in terms of magnetoresistance (MR) which is estimated using the formula $\Delta\rho/\rho_0 = (\rho_\mathrm{H}-\rho_0)/\rho_0$, where $\rho_\mathrm{H}$ is the resistivity in the presence of magnetic field. Figure \ref{fig:res}(d) shows the temperature variation of the MR percentage from 2-390 K. Overall, MR(\textit{T}) clearly depicts the magnetic transitions at $T_\mathrm{N} \sim$ 350 K and $T_\mathrm{C} \sim$ 115 K. Interestingly, MR is positive below 50 K, i.e. in the \textit{T} region-I. Low, yet negative, MR values at the transition temperatures reinstate the field-induced ferrimagnetic transition from a weak-antiferromagnetic state. The positive MR below 40 K may indicate increased magnetic-entropy. Figures \ref{fig:res}(e) and \ref{fig:res}(f) show the field-dependence of MR at certain representative temperatures in \textit{T}-regions I, II, III, and IV. At 2 K, MR is linear with \textit{H} exhibiting positive slope. At 40 K, MR is nearly zero and almost independent of \textit{H}. It is negative at all other temperatures with a changing slopes at critical fields. The temperature variation of these critical fields is in agreement with those observed in magnetization isotherms.

\begin{figure}
	\centering
	\includegraphics[width=\linewidth,height=1.5\linewidth]{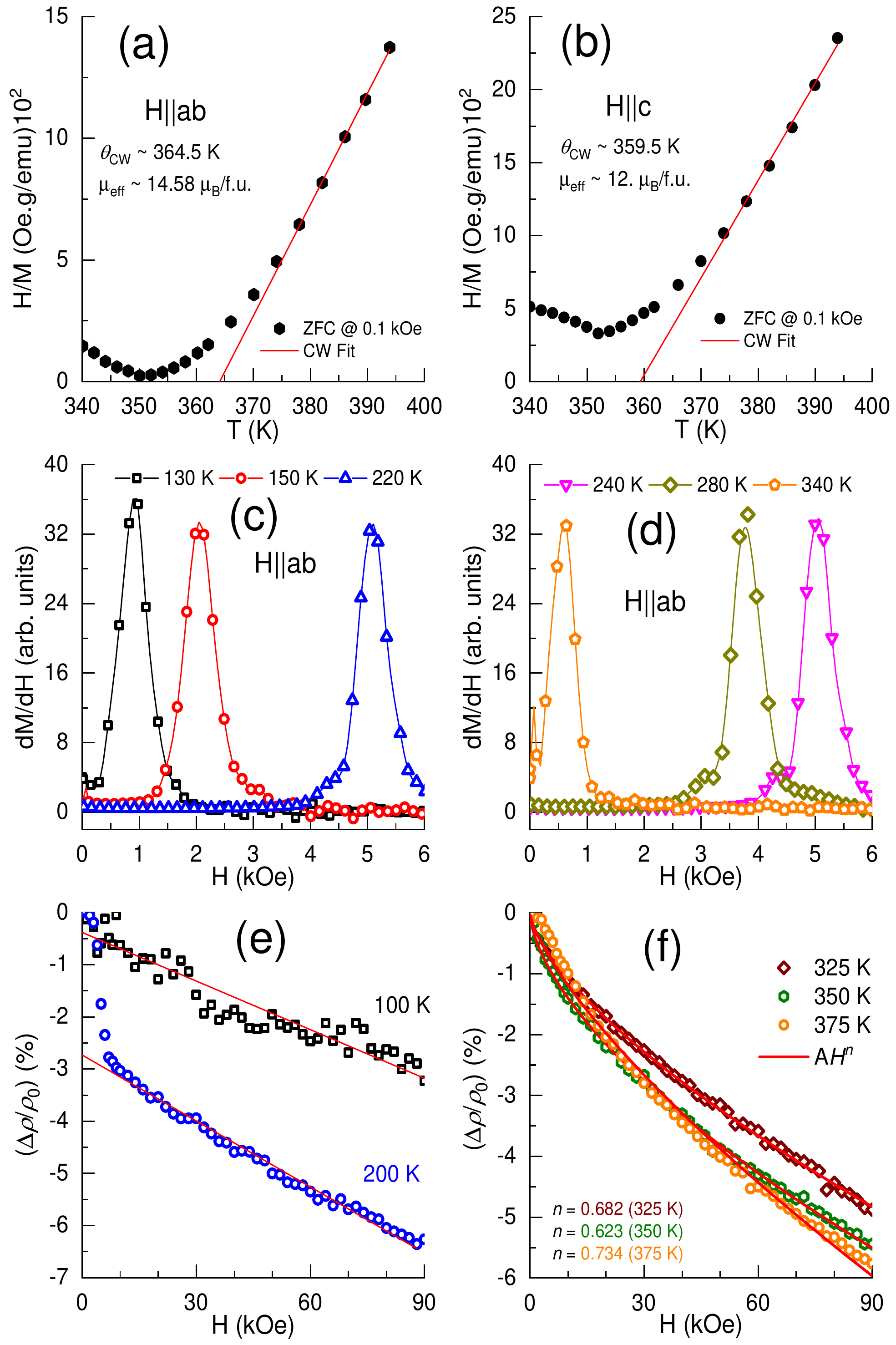}
	\caption{Curie-Weiss fits 0.1 kOe data of Er$_{0.5}$Ho$_{0.5}$Mn$_6$Sn$_6$ when (a) \textit{H}$\parallel$ab and (b) \textit{H}$\parallel$\textit{c}, along with the resulting fit parameters. Positive $\theta_\mathrm{CW}$ indicates dominant ferrimagnetic correlations. (c-d) The differential susceptibility as a function of \textit{H}, indicating crossover of AFM to FiM state across a magnetic field-region $\delta H$. (e-f). Linear fits of negative MR at 100 and 200 K above the critical fields. The deviation from a linear dependence near and above $T_\mathrm{N}$ is identified by a power-dependence ($\propto H^n$) with \textit{n} = 0.682 (at 325 K), 0.623 (at 350 K) and 0.734 (at 375 K).}
	\label{fig:CW}
\end{figure}

\begin{figure}
	\centering
	\includegraphics[width=0.8\linewidth,height=1.35\linewidth]{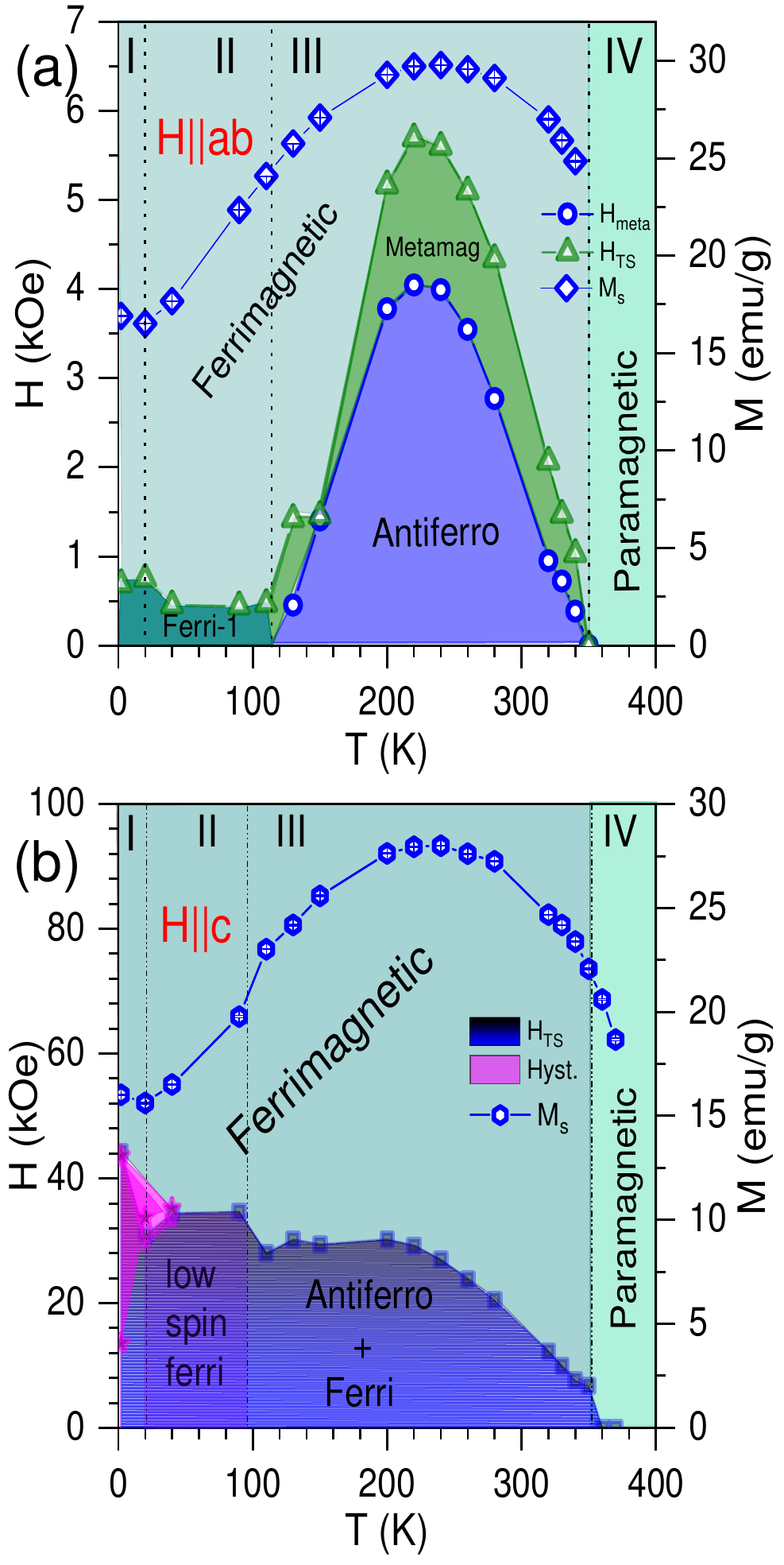}
	\caption{Temperature-field \textit{T-H} phase diagrams of Er$_{0.5}$Ho$_{0.5}$Mn$_{6}$Sn$_{6}$, constructed based on the temperature variation of the technical saturation field $H_\mathrm{TS}$ and critical fields of field-induced phase transitions. (a) \textit{H}$\parallel$\textit{ab}-plane. Region-I: a weak spin reorientation phase which quickly turns in to ferrimagnetic state in fields of 1 kOe. Region-II: zero-field ferrimagnetically correlated phase is tuned into the field-influenced ferrimagnetic saturation state. Region-III: field-induces a ferrimagnetic saturation through an intermediate spin-flop quasi-metamagnetic transition in a weakly coupled R and Mn sublattices with antiferromagnetic correlations. Region-IV represents paramagnetic state. (b) \textit{H}$\parallel$\textit{c}: Region-I: the sample exhibits a narrow field-induced hysteresis by eventually entering a ferrimagnetic state. Region-II: $H_\mathrm{TS}$ is merely independent of the temperature. Region-III: The coexistence of AFM and FiM correlations of R and Mn sublattices requires high fields of the range of 30 kOe to induce ferrimagnetic-like saturation.} 
	\label{fig:PD}
\end{figure}

\section{Discussion}

The Curie-Weiss analysis, as shown in figures \ref{fig:CW}(a) and \ref{fig:CW}(b), of inverse susceptibility $\chi^{-1}_{dc}$ ($\equiv H/M$) above $T_\mathrm{N}$, is carried out using the formula $\chi^{-1}$=$(T-\theta_\mathrm{CW})/C$ where \textit{C} is Curie constant. The Weiss temperature $\theta_\mathrm{CW}$ hints about the effective magnetic interactions in the system below the magnetic transition temperature. The fits reveal $\theta_\mathrm{CW} \sim$ 364.5 K in \textit{H}$\parallel$\textit{ab} and $\sim$ 359.5 K in \textit{H}$\parallel$\textit{c}. An effective magnetic moment $\mu_\mathrm{eff}$ (=$\sqrt{3k_\mathrm{B}C/N_\mathrm{A}}$) is found to be about 14.58 $\mu_\mathrm{B}$/f.u. in \textit{H}$\parallel$\textit{ab} and 12 $\mu_\mathrm{B}$/f.u. in \textit{H}$\parallel$\textit{c}, where $k_\mathrm{B}$ and $N_\mathrm{A}$ are Boltzmann constant and  Avogadro's number. It can be noticed that, along the \textit{ab}-basal plane, $\mu_\mathrm{eff}$ is larger and  $\theta_\mathrm{CW}$ is about 4 K higher as compared to those of in the \textit{c}-axis. For Er$_{0.5}$Ho$_{0.5}$Mn$_6$Sn$_6$, $\theta_\mathrm{CW}$ value in \textit{H}$\parallel$\textit{c} and \textit{H}$\parallel$\textit{ab} configurations is positive. In principle, it indicates the ferromagnetic/ferrimagnetic correlations below the magnetic transition temperature. However, the observed transitions are antiferromagnetic (below 350 K) and ferrimagnetic (below 114 K). This is contrary to the prediction by $\theta_\mathrm{CW}$. Therefore, the observed positive $\theta_\mathrm{CW}$ of a high-temperature antiferromagnet indicates weak AFM correlations. As a result, the compound exhibits ferrimagnetic behavior at low temperatures below 114 K. In addition, this weakly correlated antiferromagnetic state shows a tendency of becoming a ferrimagnetic state in the presence of large enough magnetic fields, as evident from the magnetic isotherms measured between $T_\mathrm{C}$ and $T_\mathrm{N}$ (region-III).

The anisotropic magnetic properties of Er$_{0.5}$Ho$_{0.5}$Mn$_6$Sn$_6$, governed by R-R and Mn-Mn (in the \textit{ab}-basal plane) and R-Mn (along \textit{c}-axis) interactions are evident from figures \ref{fig:mag_along_c} and \ref{fig:mag_along_ab}. Here, we discuss and illustrate the magnetic and electrical properties (along with their manifestation under the influence of magnetic field) of the present alloy through temperature-field \textit{T-H} phase diagrams in \textit{H}$\parallel$\textit{ab} and \textit{H}$\parallel$\textit{c} directions. The phase diagrams are constructed based on the temperature variation of different critical fields, technical saturation field and the spontaneous magnetic moment, $M_\mathrm{s}$. $M_\mathrm{s}$ is the \textit{y}-intercept of a linearly extrapolated high field magnetization as a function of \textit{H}. Nevertheless, the phase diagrams shown in figure \ref{fig:PD} confirm anisotropic magnetic behavior of the compound. 

\textit{H}$\parallel$\textit{ab}: The field manifested magnetic phases when \textit{H}$\parallel$\textit{ab} is illustrated in \textit{T-H} phase diagram, as shown in figure \ref{fig:PD}(a). Region-I and II: low-spin FiM state turns to high-spin FiM state in an applied field of about 1 kOe. Region-III: As shown in the crystal structure (figure\ref{fig:xrd_structure}(a)), in \textit{ab}-plane two Mn atomic chains (with three-atoms; one each on four faces and one at the center) are on either sides of R-atomic chain (with one-atom; each located on four edges). When \textit{H}$\parallel$\textit{ab}-plane, Mn and R-atomic chains contribute independently. R-Mn inter-planar interaction dominated AFM state transforms to FiM via metamagnetic region in a maximum field of about 5.7 kOe. A careful look into the in-field conversion of AFM to another high-magnetic moment state (FM or FiM) is required with respect to the generalization of metamagnetic transitions. In principle, a metamagnetic transition occurs when AFM spin configuration ($\uparrow\downarrow$) suddenly flips over to FM spin configuration ($\uparrow\uparrow$) at a single critical field, leading to a sharp first order transition. Such an abrupt flipping is feasible when the strong magneto-crystalline anisotropic energy is overcome by the applied \textit{H}. On the other hand, the AFM to high-magnetic moment FM/FiM transition is guided by a gradual spin-flop, instead of spin-flip, process, in the case of weak-magneto-crystalline anisotropy \cite{Solid_State_Physics_Ashcroft}, leading to quasi-metamagnetic transition without hysteresis/first-order characteristics \cite{Quasi_Meta_1_Samatham_JMMM_2017,Quasi_Meta_2_PhysRevB_2022,Quasi_Meta_3_JMMM_1999,Quasi_Meta_4_JMMM_2017,Quasi_Meta_5_PhysRevB_2017,Quasi_Meta_6_PhysRevB_2017,Quasi_Meta_7_JMMM_2019}. \textit{H}$\parallel$\textit{ab}-plane, \textit{M-H} curves in region-III exhibit gradual/continuous transition across $H_\mathrm{meta}$ with no hysteresis. Figure \ref{fig:CW}(c) shows differential susceptibility of 130, 150 and 220 K as a function of \textit{H}. It is noticed that a metamagnetic transition is distributed over a certain field range, indicating its quasi-behavior. Weak-AFM correlations inferred from large and positive $\theta_\mathrm{CW}$, small $H_\mathrm{meta}$ of about 4 kOe, no field-cycling hysteresis and relatively broad distribution of transition field suggests quasi spin-flop metamagnetic transition from zero-field AFM to field-induced FiM, along \textit{ab}-plane. This is in contrast to the coexisting mixed-magnetic phases along the \textit{c}-axis.

\textit{H}$\parallel$\textit{c}: Figure \ref{fig:PD}(b) is \textit{T-H} phase diagram of Er$_{0.5}$Ho$_{0.5}$Mn$_6$Sn$_6$ along \textit{c}-axis, which illustrates the field-induced magnetic behavior of the compound in four different temperature regions; I-$T<T_\mathrm{tr}$, II-$T_\mathrm{tr} \le T \le T_\mathrm{C}$, III-$T_\mathrm{C} \le T \le T_\mathrm{N}$ and $T>T_\mathrm{N}$. It is evident that the order parameter $M_\mathrm{s}$ is zero in the paramagnetic (disorder) state and varies smoothly below the ordering temperature, indicating a second order transition from paramagnetic to a field-induced ferrimagnetic state. $H_\mathrm{TS}$ is estimated as the field at which the sharply rising magnetization begins to saturate in principle. It decreases with temperature as $T$ approaches $T_\mathrm{C}$.  Below $T_\mathrm{N}$, the magnetic behavior along \textit{c}-axis is governed by the possible symmetric interactions of R and Mn atomic chains; i) majority R$_\mathrm{edge}$-Mn$_\mathrm{face}$-R$_\mathrm{edge}$, Mn$_\mathrm{face1}$-R$_\mathrm{edge}$-Mn$_\mathrm{face2}$ and diagonally R$_\mathrm{edge}$-Mn$_\mathrm{center}$-R$_\mathrm{edge}$ interactions with $d_\mathrm{R-Mn}$ = 3.1866(7) \AA, ii) Mn$_\mathrm{face1}$-Mn$_\mathrm{center}$-Mn$_\mathrm{face2}$ symmetric interaction with $d_\mathrm{Mn-Mn}$ = 2.7597(6) \AA~and iii) minor R$_\mathrm{edge}$-Mn$_\mathrm{center}$-R$_\mathrm{edge}$ interactions (diagonally) with $d_\mathrm{Mn-R}$ = 5.2722(9) \AA~distances. In region-I and II, the zero-field low-spin ferrimagnetic configuration state transforms into a field-induced high-spin ferrimagnetic state. This has led to an opening of hysteresis, which ceases to exist at $T>$ 40 K. Confining to region III, $H_\mathrm{TS}$-boundary separates the field-induced ferrimagnetic phase from the zero-field \textit{c}-axis magnetic phase. The experimental $\mu_\mathrm{eff}$ is smaller than the theoretically estimated effective contribution $\mu^\mathrm{Th}_\mathrm{eff} \sim$ 15.7 $\mu_\mathrm{B}$/f.u. $\mu^\mathrm{Th}_\mathrm{eff} = \sqrt{1\times\mu_\mathrm{eff,R}^{2}+6\times\mu_\mathrm{eff,Mn}^{2}}$ where $\mu_\mathrm{eff,R} = \sqrt{0.5\mu_\mathrm{eff,Er}^{2}+0.5\mu_\mathrm{eff,Ho}^{2}}$, where $\mu_\mathrm{eff,Er}$ = 9.59, $\mu_\mathrm{eff,Ho}$ = 10.6 and $\mu_\mathrm{eff,Mn}$ = 4.91 $\mu_\mathrm{B}$ \cite{Solid_State_Physics_Kittel}. The estimated effective magnetic moments apply to paramagnetic region. An AFM undergoes either a field-induced metamagnetic phase transition in sufficient magnetic fields or remains antiferromagnetic. AFM to FM/FiM conversion without a metamagnetic transition is rare unless there exist mixed phases of different magnetic behavior. In region-III, a linearly increasing \textit{M-H} and a high-field FiM suggest the coexistence/mixed phases of FiM and dominant AFM, limited to $H_\mathrm{TS}$ boundary. Eventually, in region-IV i.e., in the paramagnetic phase, $H_\mathrm{TS}$ becomes zero.

The electrical resistivity $\rho$-\textit{T} of Er$_{0.5}$Ho$_{0.5}$Mn$_6$Sn$_6$ exhibits change in slopes at $T_\mathrm{N}$ and $T_\mathrm{C}$ as shown in figure \ref{fig:res}(a). A fit of zero-field resistivity $(\rho-\rho_0)$ fits well by using $T^2$, a contribution from coherent electron-magnon (\textit{e-mag}) scattering and $T^{3/2}$ \cite{Res_Temp_Depend_2_Kettler_1989_PhysRevB,Res_Temp_Depend_1_Paja_1990_PSSB}, arising from incoherent \textit{e-mag} scattering, terms. However, $T^2$ dependence at low temperatures also represents the \textit{e-e} interaction \cite{e_e_1_interaction_Mills_1966,e_e_2_interaction_Volkenshtein_1973,e_e_3_interaction_Maldague_1979}, suggesting nearly Fermi liquid behavior \cite{Fermi_liquid_Stewart_RevModPhys_1984} as inferred from the sizable \textit{A} = (24.7 $\pm$ 0.6)$\times10^{-3}$ $\mu \Omega$cm$T^{-2}$. In regions II, III and IV, MR is negative and its absolute magnitude is small with a maximum of 6\%, refer to figures \ref{fig:res}(d)-(f). Such low and negative MR is attributed to arise from the quenching of spin-disorder scattering/spin fluctuations by \textit{H}, thereby reducing the spin-scattering \cite{LMR_CDW_3_PNAS_2019}. Figure \ref{fig:CW}(e) shows the representative \textit{MR-H} data at 100 and 200 K. Classically, in magnetic/non-magnetic metals, quadratic dependence (at low fields) and saturation (at high fields) of MR is reported to be because of uncompensated electron-hole densities \cite{MR_Classical_Spin_disorder_Ziman,MR_Classical_Spin_disorder_Pippard,MR_Classical_Spin_disorder_Ziman}. Such a possibility is ruled out in the present sample since MR shows non-quadratic field dependence, as shown in figures \ref{fig:CW}(e) and \ref{fig:CW}(f). MR varies linearly with \textit{H} above the metamagnetic critical field $H_\mathrm{meta}$. However, $H_\mathrm{meta}$ is zero at 100 K. Interestingly, in region-I, MR is positive, large (as compared to other magnetic regions) and linear, refer to figure \ref{fig:res}(e). A linearly depending positive MR (LMR) in low fields is earlier reported in some of the charge density wave materials \cite{LMR_CDW_1_PhysRevB_2017,LMR_CDW_2_PNAS_2018,LMR_CDW_3_PNAS_2019,LMR_CDW_4_2D_Mater_2023} and non-magnetic silver chalcogenides \cite{LMR_Xu_Nature_1997} and topological insulators \cite{LMR_Topological_4_PhysRevB_2012,LMR_Topological_1_AIP_Advances_2017,LMR_Topological_2_APL_2018,LMR_Topological_3_JPCM_2018}. Linear MR finds the quantum route when electron orbits are confined to the individual quantum levels, leading to quantum magnetoresistance (QMR) described using the generic relation $\rho \propto N_i/en^2H$ \cite{LMR_Abrikosov_2_1998,LMR_Abrikosov_1_2000,LMR_Hu_Nature_Mater_2008}. Further, as \textit{T} $\rightarrow$ $T_\mathrm{N}$, an almost square-root field dependence of MR is evident. The possibility of QMR in layered kagome magnetic metal Er$_{0.5}$Ho$_{0.5}$Mn$_6$Sn$_6$ at low temperature and high magnetic fields is beyond the scope of the current discussion and will be investigated separately in the future. In addition, it is worth noticing that negative MR at $T_\mathrm{N}$ reinstates the ferrimagnetic state as an actual magnetic ground state of compound under study in region-III, as is also suggested by positive $\theta_\mathrm{CW}$. In addition, oppositely strongly correlated R-Mn spin chains caused mixed magnetic phases in zero-fields. Nevertheless, the destabilization of zero field (ZF) antiferromagnetic correlations in magnetic fields should, in principle, lead to a metamagnetic transition to a field-induced FiM state. In such cases, $\rho(H)$ exhibits an abrupt drop at a single critical field ($H_\mathrm{meta}$), showing a first-order phase transition. However, the experimental results of resistivity (\textit{H}$\parallel$\textit{c}) and magnetization (\textit{H}$\parallel$\textit{ab}-plane and \textit{c}-axis) of the present compound are in support of smooth conversion of ZF-(AFM/AFM-FiM) to FiM state, thereby suggesting a quasi-metamagnetic non-first-order phase transition.

\section{Summary}

The effect of magnetic field on the anisotropic magnetic properties of a layered kagome metal Er$_{0.5}$Ho$_{0.5}$Mn$_6$Sn$_6$ with dual transitions ($T_\mathrm{N} \sim$ 350 K and $T_\mathrm{C} \sim$ 114 K) is investigated using magnetization measurements on a high quality single crystal. The temperature evolution of distinct magnetic properties along with their manifestation in fields when \textit{H}$\parallel$\textit{ab} and \textit{H}$\parallel$\textit{c} is depicted through \textit{T-H} phase diagrams. In \textit{H}$\parallel$\textit{ab}-plane, a weak-antiferromagnetic phase between $T_\mathrm{N}$ and $T_\mathrm{C}$ undergoes a spin-flop quasi-metamagnetic phase transition to a ferrimagnetic ground state under the influence of small magnetic fields of about 4 kOe. In \textit{H}$\parallel$\textit{c}, the strong coexistence of ferri and antiferromagnetic phases requires high magnetic fields to induce a complete ferrimagnetic phase without first-order phase transition (a field-induced hysteresis). Such coexistence caused a field-induced narrow hysteresis at low temperatures. The temperature dependent electrical resistivity suggests metallic nature with Fermi liquid behavior at low temperatures. Negative and small magnetoresistance is attributed to the magnetic field quenching of spin-disorder scattering. The relatively large, positive and linear magnetoresistance suggest its routes in quantum effects and needs further exploration. Essentially, the current study stimulates interest to investigate substitution induced i) unusual electrical and Hall transport properties and manifested magnetic and physical properties of layered kagome magnetic metals for possible novel and exotic behavior, despite negligible chemical pressure.

\section*{Acknowledgments}
This work was performed at the State University of New York (SUNY), Buffalo State University, and supported by the National Science Foundation, Launching Early-Career Academic Pathways in the Mathematical and Physical Sciences (LEAPS-MPS) program under Award No. DMR-2213412. JC and NK acknowledge financial support from the Office of Undergraduate Research, SUNY, Buffalo State University. SSS acknowledges the financial support through Core Research Grant (CRG/2022/0007993).

\bibliography{refs_ehms}

\end{document}